# First-Principles-Derived Effective Mass Approximation for the Improved Description of Quantum Nanostructures

**Hyeonwoo Yeo[1], Jun Seong Lee[1], Muhammad Ejaz Khan[1†], Hyo Seok Kim[1], Duk Young Jeon[2], and Yong-Hoon Kim[1]***

[1] School of Electrical Engineering, Korea Advanced Institute of Science and Technology (KAIST), 291 Daehak-ro, Yuseong-gu, Daejeon 34141, Korea.
[2] Department of Materials Science and Engineering, Korea Advanced Institute of Science and Technology (KAIST), 291 Daehak-ro, Yuseong-gu, Daejeon 34141, Korea.

[†] Current affiliation: Department of Computer Engineering, National University of Technology, I-12, Islamabad 44000, Pakistan.

E-mail: y.h.kim@kaist.ac.kr

## Abstract

The effective mass approximation (EMA) could be an efficient method for the computational study of semiconductor nanostructures with sizes too large to be handled by first-principles calculations, but the scheme to accurately and reliably introduce EMA parameters for given nanostructures remains to be devised. Herein, we report on an EMA approach based on first-principles-derived data, which enables accurate predictions of the optoelectronic properties of quantum nanostructures. For the CdS/ZnS core/shell quantum rods, for which we recently reported its experimental synthesis, we first carry out density functional theory (DFT) calculations for an infinite nanowire to obtain the nanoscopic dielectric constant, effective mass, and Kohn-Sham potential. The DFT-derived data are then transferred to the finite nanorod cases to set up the EMA equations, from which we estimate the photoluminescence (PL) characteristics. Compared with the corresponding method based on bulk EMA parameters and abrupt potential, we confirm that our EMA approach more accurately describes the PL properties of nanorods. We find that, in agreement with the experimentally observed trends, the optical gap of nanorods is roughly determined by the nanorod diameter and the PL intensity is reduced with increasing the nanorod length. The developed methodology is additionally applied to CdSe nanoplatelets, where reliable experimental data became recently available. Here, we again obtain excellent agreements between calculated and measured optical gap values, confirming the generality of our approach. It is finally shown that the abrupt confinement potential approximation most adversely affects the accuracy of EMA simulations.

## 1. Introduction

With the advancement in nanofabrication and synthesis techniques, it is now possible to prepare semiconductor nanostructures with various sizes and shapes to tune their electronic and optical properties [1, 2]. In these nanostructures that experience quantum and dielectric confinement effects, one can engineer the excitons to acquire features that are beneficial for various device applications such as light-emitting diode (LED), photosensor, solar cell, solar fuels production, and biological labelling [3]. Particularly, their optical properties can be further modulated by inducing various external stimuli such as the electric and magnetic fields, which particularly make them promising candidates for display applications.

In the characterization and design of the semiconductor nanostructures, computer simulations have been playing an important role. Here, in principle, first-principles schemes such as density functional theory (DFT) and many-body or quantum Monte Carlo simulations performed on top of DFT would be desirable [4]. In practice, however, DFT and DFT-based higher-level calculations that require very large computational resources are often too demanding or even impossible to be applied to nanostructures of realistic sizes. Accordingly, approximate large-scale methods such as the effective mass approximation (EMA) and tight binding techniques are still routinely employed for the study and design of large complex semiconductor nanostructures [5, 6]. However, employing the effective mass and dielectric constant derived from the bulk crystal, the EMA approach often fails to produce accurate and reliable results for nanostructures experiencing quantum and dielectric confinement effects.

In this work, in view of the optoelectronic device applications based on semiconductor nanostructures, we extend the EMA simulator based on our grid-based Object-Oriented Real-space Engine (OORE; which means "thunder" in Korean) for electronic structure calculations [7-12] by employing the EMA parameters generated from first-principles calculations on reference nanostructures. Specifically, from the reference DFT calculations, we extract the nanoscopic dielectric constant, electron and hole effective masses, and additionally the Kohn-Sham (KS) potential. The envelope function of the atomistic KS potential then allows us to define an accurate EMA potential in an unambiguous manner. While our approach should be generally applicable to finite-size quantum nanostructures including zero-dimensional (0D) quantum dots, one-dimensional (1D) quantum rods, and two-dimensional (2D) nanoplatelets, we here focus on the optical properties of semiconducting nanorods and nanoplatelets that have recently seen significant experimental advances in their synthesis. Specifically, for the 1D CdS/ZnS core/shell nanorods [13] and 2D CdSe nanoplatelets [14-17], we will show that our DFT-derived EMA approach provides the optical gaps in good agreement with the experimentally measured data. For the CdS/ZnS nanorods, e.g., it will be shown that the optical gap of nanorods is mainly determined by the nanorod diameter and that the photoluminescence (PL) intensity is reduced as the nanorod length is increased. We also individually estimate the effects of the bulk effective masse, bulk dielectric constant, and abrupt confinement potential approximations, and conclude that the usage of an abrupt confinement potential most adversely affects the computational accuracy.

## 2. Computational methods

### 2.1 DFT calculations

For the 1D CdS core-only and CdS/ZnS core/shell nanowires infinitely extended along the *z*-axis and 2D CdSe nanosheets or slabs infinitely extended along the *xy*-directions, we performed for their unit cell models DFT calculations within the local density approximation (LDA) exchange-correlation functional [18]. DFT calculations were performed with the VASP package, in which the core electrons are handled by the projector augmented wave method [19]. The plane-wave basis set with a kinetic energy cutoff of 400 eV and the self-consistency cycle energy criterion of $10^{-4}$ eV were adopted. To avoid artificial interactions with the neighboring images within the periodic boundary condition, a vacuum space of more than 20 Å was inserted along the *xy* directions perpendicular to the 1D nanowire axis and along the *z* axis of the 2D slab. The Brillouin zone was sampled with a $1 \times 1 \times 10$ Monkhorst-Pack grid for CdS/ZnS nanowires and a $9 \times 9 \times 1$ Monkhorst-Pack grid for CdSe nanosheets. For the nanowire case, the edge states of the $(10\bar{1}0)$ surfaces were passivated by pseudo-hydrogen atoms. Specifically, the Cd or Zn dangling bonds were passivated with the pseudo-hydrogen atom with nuclear charge $Z = 1.5$ electrons, and each S dangling bond was passivated by the pseudo-hydrogen atom with $Z = 0.5$ electrons [20]. For the nanoplatelet case, the (001) surfaces were passivated with acetate ligands. The dielectric constants of nanostructures were calculated using the optical dielectric function calculation module available within VASP.

### 2.2 EMA calculations

To assess the electronic structures and PL intensities of nanorods with different lengths and diameters, we performed EMA calculations using our grid-based OORE code [7-12]. It utilizes the higher-order finite-difference expansions of the Laplacian operator [21, 22],

$$\frac{d^2}{dx^2} f(x) = \sum_{j=-N}^{N} C_j f(x + jh) + O(h^{2N+2}), \qquad (1)$$

where $h$ is the grid spacing and $C_j$ are the finite-difference coefficients, and the multigrid iterative minimization schemes for the solutions of Schrödinger and Poisson equations. The OORE framework includes general tools to carry out grid-based first-principles DFT calculations [7], and by simply replacing pseudopotentials by EMA potentials one can perform large-scale 3D EMA calculations (OOREQD) with even including the exact-exchange electron interaction [8, 9]. The key features and further developments relevant for the calculation of optical properties of semiconductor nanostructures within the newly-developed DFT-based EMA scheme will be presented in Sec. 3.

## 3. Formulation of the DFT-derived EMA approach

### 3.1 General strategy

Figure 1 graphically summarize the strategy of the DFT-based EMA scheme proposed in this work. For the "ideal" low-dimensional nanostructures, we first carry out DFT calculations and obtain the electron/hole effective mass $m^*_{e/h}$ and dielectric constant $\epsilon$, and KS potential $v^{KS}$. Note that the effective mass and dielectric constant together set the length scale as $a^*_{e/h} = \epsilon/m^*_{e/h}$ and the energy scale as $Ry^*_{e/,h} = m^*_{e/h}/\epsilon^2$. Here, we define the "ideal" systems as the nanostructures that are infinitely extended along the non-confined directions. For example, for the finite quasi-1D nanorods and quasi-2D nanoplatelets, we consider 1D nanowires and 2D nanosheets, respectively, with the periodic boundary condition (PBC) along the *z*- and *xy*-directions, respectively. For the zero-dimensional quantum dots, we could adopt a reasonably-sized quantum dot and employ the nanostructure-derived dielectric constant and KS potential in combination with the bulk effective mass. Throughout Sec. 3, we will take the CdS/ZnS nanorod case as the representative example and discuss in more detail the procedure of systematically employing nanoscopic EMA parameters derived from DFT calculations. Note that, in view of the LED applications, the type-I band alignment across the core/shell nanorod will be assumed.

### *3.2 The effective mass approximation formulation*

We solve within the isotropic EMA framework the conduction band edge (electron) Schrödinger equation for the electron wavefunction $\psi_e$ and energy $E_e$,

$$\left[-\frac{\hbar^2}{2m_e^*}\boldsymbol{\nabla}^2 + v^{KS}_{eff,e}(\boldsymbol{r}_e)\right]\psi_e(\boldsymbol{r}_e) = E_e\psi_e(\boldsymbol{r}_e), \quad (2)$$

and separately the valence band edge (hole) Schrödinger for the hole wavefunction $\psi_h$ and energy $E_h$,

$$\left[-\frac{\hbar^2}{2m_h^*}\boldsymbol{\nabla}^2 + v^{KS}_{eff,h}(\boldsymbol{r}_h)\right]\psi_h(\boldsymbol{r}_h) = -E_h\psi_h(\boldsymbol{r}_h), \quad (3)$$

where $\hbar$ is the reduced Planck's constant, $m_e^*$ is the electron effective mass, and $m_h^*$ is the hole effective mass.

We emphasize that the key development in this work is the adoption of the EMA parameters derived from first-principles calculations performed on the representative model nanostructures. Importantly, in addition to the dielectric constant and effective masses, we introduce the effective potential $v^{KS}_{eff,e/h}$ from the reference DFT calculations. At the fundamental level, it was argued that the "exact" DFT KS equations for $N$ electrons can be characterized as the Dyson equation for $N-1$ electrons, so unoccupied orbitals obtained in the KS calculations should physically describe number-conserving optical excitations of the $N$-electron system [7, 23-26]. While we adopt $v^{KS}_{eff,e} = v^{KS}_{eff,h} \equiv v^{KS}_{eff}$ with this physical nature of the KS potential in mind, given that we start from LDA DFT calculations contaminated by self-interaction errors, we heuristically regard equations (2) and (3) as quasi-particle equations [4] and determine the expressions for quasiparticle and optical gaps of quantum nanostructures within EMA as described below.

Once the hole and electron Schrödinger equations are solved, we estimated the exciton transition energy or optical gap $E_g^{opt}$,

$$E_g^{opt} = E_g^{qp} - E_X, \quad (4)$$

by calculating the band edge transition energy or quasiparticle gap $E_g^{qp}$ according to

$$E_g^{qp} = E^{qp}_{g,bulk} + E_e - E_h. \quad (5)$$

The CdS bulk quasiparticle gap $E^{qp}_{g,\mathrm{bulk}}$ was obtained by adding the experimentally reported bulk optical bandgap value $E^{opt}_{g,\mathrm{bulk}}$ of 2.42 eV [27] to the bulk exciton binding energy of 0.026 eV calculated according to

$$E_{X,bulk} = \frac{\mu e^4}{32\pi^2\hbar^2\epsilon_r^2\epsilon_0^2}, \quad (6)$$

where $\epsilon_r$ and $\epsilon_0$ are static bulk dielectric constant and vacuum permittivity, and $\mu$ is the reduced effective mass,

$$\frac{1}{\mu} = \frac{1}{m_e^*} + \frac{1}{m_h^*}. \quad (7)$$

For the nanorods, we then calculated the exciton binding energy $E_X$ using the expression,

$$E_X = \iint \frac{|\psi_h(\vec{r}_h)|^2|\psi_e(\vec{r}_e)|^2}{\epsilon\,|\vec{r}_h - \vec{r}_e|} d\vec{r}_e d\vec{r}_h. \quad (8)$$

Finally, the oscillator strength for the electron-hole band edge exciton transition was calculated according to [28, 29]

$$O_{eh} = \frac{2m_e^*\,\omega}{\hbar}|\langle\psi_e|z|\psi_h\rangle|^2, \quad (9)$$

where $O_{eh}$ presents the oscillator strength for the electron transition from $\psi_e$ at $E_e$ to $\psi_h$ at $E_h$.

### *3.3 Effective mass and dielectric constant from DFT calculation*

For the descriptions of the CdS/ZnS core/shell nanorods, for which we recently reported relevant experimental data [13], we first carried out PBC DFT calculations on the corresponding 1D nanowires (see Ref. [13] for the details). Denoting the nanowire model with $m$ CdS and $n$ ZnS layers as $(CdS)_m(ZnS)_n$, we show in figure 2(a) the $(CdS)_3(ZnS)_2$ core/shell nanowires with diameter of ~ 3.5 nm (excluding passivating pseudo-hydrogen atoms) optimized within DFT. From these models, we derived the electron (hole) effective mass $m_e^*$ ($m_h^*$) and dielectric constant $\epsilon$.

The DFT-calculated dispersions of the conduction (top panels) and valence (bottom panels) band edges of the $(CdS)_3(ZnS)_2$ nanowire are shown in figure 2(b). Note that the significant lattice strain effect has been already reflected on the band structures through DFT geometry optimizations [13]. The procedure of extracting electron (hole) effective mass from the DFT-derived conduction (valence) band dispersion curve is also schematically described in figure 2(b), where the region of band dispersion used for effective mass fitting is marked with the shaded rectangle near the gamma (Γ) $k$-point. To obtain the electron and hole effective masses, we adopted the harmonic $E$-$k$ dispersion relation near the Γ according to

$$E(k) = E_0 \pm \frac{\hbar^2 k^2}{2m_{e,h}^*}, \tag{10}$$

where $E_0$ is the energy eigenvalue of selected conduction minimum (CBM) or valence band maximum (VBM) used for the effective mass fitting. The effective masses fitted to the CBM and VBM of the $(CdS)_3(ZnS)_2$ nanowire are presented in table 1. While the bulk CdS-derived electron and hole effective masses are 0.2 $m_0$ and 0.7 $m_0$, respectively [30, 31], the corresponding values derived by fitting equation (11) to $(CdS)_3(ZnS)_2$ band edges are 0.2 $m_0$ and 0.51 $m_0$, respectively. Namely, we determine that while the bulk electron effective mass is translated into the nanowire electron effective mass $m_e^*$, the hole effective mass $m_h^*$ is reduced by ~ 30 % through nanostructuring.

Regarding the dielectric constants of the CdS/ZnS nanowires, as summarized in table 1, we find that they are significantly decreased from the bulk CdS dielectric constant value of 8.92 [32] due to the reduced electronic screening effect [33-36]. Quantitatively, for the $(CdS)_3(ZnS)_2$ core/shell nanowire, the optical dielectric constant values along the axial and radial directions were $\epsilon_r^{zz} = 2.3$ and $\epsilon_r^{xx,yy} = 2.2$, respectively. Note the small difference between the dielectric constants along the radial and axial directions, which indicates the negligible anisotropy in the local dielectric screening environment. Accordingly, we will adopt the isotropic dielectric constant within the EMA calculations.

### *3.4 The EMA potential from DFT calculation*

As emphasized earlier, in addition to the effective mass and dielectric constant, the utilization of the KS potential $v_{KS}$ information to construct the EMA effective potentials $v_{eff}$ represents a key feature of our approach. Note that in general the confinement potential shape is a critical factor in determining the electronic and optical properties of quantum nanostructures. For example, it was theoretically suggested that the suppression of undesirable nonradiative Auger processes can be achieved by smoothing out the confinement potential or increasing the core volume[5, 6]. Accordingly, much experimental efforts were recdently devoted to understand and optimize the material gradient at the core/shell interface [37, 38].

In figures 2(c) and (d), we present the cylindrically-averaged DFT KS potentials and the corresponding EMA effective potentials obtained for the $(CdS)_3(ZnS)_2$ and $(CdS)_3(ZnS)_1$ nanowire cases, respectively. The macroscopically smooth EMA potentials were generated by obtaining the envelope functions of the KS potentials that oscillate at the atomic scale using a double filtering process with the step function as the filter function [39],

$$w(r) = \frac{1}{l}\theta\left(\frac{l}{2} - |r|\right), \tag{11}$$

Here, we chose the smoothening parameter $l \approx 7$ Å, which is approximately the radial thickness of two CdS (or ZnS) layers. This choice was made because, as discussed in Supplementary figure S1, the lattice periodicity of the hexagonal nanowire lattice geometry imposes the minimum of $l \approx 6.2$ Å. As shown in Supplementary figure S1, we confirmed that changing the $l$ value by up to about $\pm 1$ Å negligibly change the smoothened potential shape. The radially smoothened 1D EMA potential profiles were then directly projected along the boundaries of quantum rods, for which we adopted the rectangular or cylindrical shapes as shown in figures 3(a) and 3(b), respectively. To confirm the importance of the DFT-based EMA effective potential, as shown in figure 3(c), we additionally adopted an abrupt potential with the potential depth fixed to the DFT-derived EMA potential value. Based on the nature of DFT KS equations mentioned above [7, 23-26], we used the same EMA potential profile (with the opposite sign) for both hole and electron wavefunctions.

## 4. Applications of the DFT-derived EMA approach

### *4.1 CdS/ZnS nanorods*

#### *4.1.1 Comparison of wavefunctions from DFT and EMA calculations.*

To check the quality of the EMA potentials employed in our scheme, we first analyzed the radial-direction EMA electron and hole wavefunctions obtained by solving equations (3) and (4), respectively, against their DFT counterparts. In figure 4(a), we first show the CBM (top) and VBM (bottom) wavefunctions obtained from the DFT calculations performed on the $(CdS)_3(ZnS)_2$ nanowire. We next show in figure 4(b) the corresponding wavefunctions obtained from the DFT-based EMA calculations performed for a 12 nm-long $(CdS)_3(ZnS)_2$ rectangular-shape nanorod. We find that, as shown in Supplementary figure S2, very similar wavefunctions are obtained by adopting the cylindrical shape EMA potential. Additionally, in figure 4(c), we show the corresponding wavefunctions obtained with the abrupt EMA potential [figure 3(c)].

Overall, as can be expected by the comparison of the DFT and EMA potentials, we observe that the atomic scale oscillations in the DFT-derived wavefunctions [figure 4(a)] are smoothend out in the EMA envelope wavefunctions [figures 4(b) and 4(c)]. Next, comparing the EMA calculations based on the DFT-derived and abrupt EMA potentials, we can notice that our DFT-derived EMA scheme much more closely reproduces the envelope profiles of DFT wavefunctions: Both the electron and hole wavefunctions penetrate into the shell region, and particularly the electron wavefunctions exhibit more delocalized nature. On the other hand, the abrupt effective potential-based EMA method results in wavefunctions that are too strongly confined within the core region [figure 4(c)], leading us to conclude that our DFT-based EMA approach indeed represents an improvement in describing quantum nanostructures.

#### *4.1.2 Optical properties of CdS/ZnS nanorods from EMA calculations*

We now consider the energy gaps of quantum rods computed in our EMA approach. In figure 5(a), we present the quasiparticle gap $E_g^{qp}$ of the $(CdS)_3/(ZnS)_2$ nanorod with the lengths of 18 and 24 nm with the cylindrical (black triangle) and rectangular (blue square) confinement potential shapes shown in figures 3(a) and 3(b), respectively. In both cases, $E_g^{qp}$ shows negligible changes ($\lesssim$ 5 meV) with respect to the nanorod length, indicating that $E_g^{qp}$ or electron/hole eigenvalues $E_e/-E_h$ of nanorods is essentially determined by the smaller-dimension or radial-direction quantum confinement. Comparing the $E_g^{qp}$ values obtained from the cylindrical and rectangular EMA potentials, we find that the former is about 0.1 eV larger than the latter because of the slightly smaller cross section in the cylinder (for a fixed radius $r$, $\pi r^2$ rather than $4r^2$ in the rectangular rod shape). After all, due to the close correspondence between the wavefunctions and eigenvalues or $E_g^{qp}$ values from the cylindrical and rectangular shape confinement potentials, we will from now on consider only the rectangular EMA potential case.

For comparison, we also present in figure 5(a) the $E_g^{qp}$ values obtained from the EMA calculations using the bulk effective masses and dielectric constant together with the abrupt confinement potential profile (red star) shown in figure 3(c). They are smaller than those obtained from DFT-based EMA calculations by about 0.6 eV, indicating the weaker quantum confinement (due to the potential bottom area wider than that in the DFT-derived potential case) and providing an estimate of the error produced by the abrupt potential approximation.

In figure 5(b), we next show the optical gap $E_g^{opt}$ values calculated according to equation (4) by subtracting $E_X$ from $E_g^{qp}$ together with the experimentally measured $E_g^{opt}$ values (black filled circles) [13]. Within nanostructures, the quantum and dielectric confinement effects induce enhanced electron-hole interactions, increasing the exciton binding energy $E_X$. This feature of increasing $E_X$ at nano scale also results from the reduced dielectric constant or electronic screening in nanostructures compared to the bulk limit [33-36] (see section 3.3 and table 1). For example, it was experimentally observed that the $E_X$ value of CdSe significantly increase from 15 meV [40] in the bulk limit to about 240 meV in nanorods, and to about 400 meV in quantum dots [41, 42]. The $E_X$ values of CdS nanorods with diameters of 4 ~ 10 nm were reported to be in the range of 220 ~ 300 meV [41], again an order of magnitude larger than the bulk CdS $E_X$ value of 28 meV [43].

The $E_X$ values calculated from our DFT-based EMA calculations of $(CdS)_3(ZnS)_2$ according to equation (8) are in the range of 345 ~ 454 meV, which goes together with the trend of reported $E_X$ values[41]. Then, the resulting $E_g^{opt}$ values obtained within our DFT-based EMA approach (blue open squares) are in excellent quantitative agreement with the experimental data. As a reference, we also calculated the $E_X$ value using the expression for a spherical quantum dot with the radius $R$ [44-48],

$$E_X = -\frac{1.786\, e^2}{\epsilon R}. \quad (12)$$

For the $(CdS)_3$ case with the core diameter of about 2.3 nm, this leads to the $E_X$ = 588 meV. This value that is larger than the nanorod $E_X$ values (345 ~ 454 meV) is reasonable in that the quantum confinement should be further increased as one going from the nanorod geometry to the quantum dot limit.

On the other hand, using the abrupt confinement potential and bulk effective masses and dielectric constant, we obtained the $E_X$ values of 51 ~ 77 meV and the $E_g^{opt}$ values smaller than the experimental ones by ~ 200 meV (red stars). This $E_X$ value is apparently a significant underestimate of the experimental values [41, 43], indicating the shortcoming of the conventional EMA approach and quantifying the improvement that can be achieved in our EMA scheme.

With the electron and hole wavefunctions obtained from DFT-based EMA calculations, we also evaluated the oscillator strength according to equation (9) and estimated the PL intensity for the CBM-to-VBM transition. In figure 5(c), we show the PL intensities of the $(CdS)_3/(ZnS)_2$ nanorods in the cases of nanorod lengths 16 nm and 24 nm. For the 24 nm nanorod case, we obtained the PL peak position of 431 nm, which is in excellent agreement with the experimental value [13]. For the shorter 16 nm nanorod, due to the increased axial direction quantum confinement, we obtained an about 30% increase of the PL intensity. The trend of enhanced PL, or the higher probability of electron-hole recombination, in shorter nanorods is again in good agreement with experimental data [13].

### 4.2 CdSe nanoplatelets

#### 4.2.1 EMA calculations of the optical gaps of CdSe nanoplatelets

For the CdS/ZnS nanorod case, in spite of the high level of agreement between theory and experiment, the two experimental $E_g^{qp}$ data points are admittedly very limited [13]. To further confirm the generality and accuracy of the developed formalism, we next considered the CdSe nanoplatelet case, for which accurate $E_X$ and $E_g^{opt}$ experimental data have been reported in recent years [14-17]. In figure 6(a), we show the unit cell of a reference 2D CdSe nanosheet with the thickness of 4.5 monolayers and passivated with acetate ligands. Remind that, as mentioned in section 3.1, we adopt for DFT calculations the ideal reference nanostructures that preserve the infinite periodicity along the free directions. Thus, for the EMA simulation of CdSe nanoplatelets, we adopted the corresponding 2D CdSe slab and performed DFT calculations within the PBC along the *xy*-directions. In the same manner as the CdS/ZnS nanorod case, we then extracted the nanoscopic electron/hole effective masses $m_{e/h}^*$ (see Supplementary figure S3 for calcualted band structures) and dielectric constant $\epsilon$, and summarized the EMA parameters in table 2. Another important ingredient of our DFT-derived EMA scheme was the extraction of the EMA confinement potential from the reference DFT potential $v^{KS}$. We show in figure 6(b) the macroscopically smoothed EMA effective potential together with the original KS potential. In carrying out the double filtering procedure, by considering the periodicity of the lattice along the *z*-direction, we adopted the smoothening parameter $l \approx 3$ Å [see equation (11)].

With the DFT-derived EMA parameters, we next carried out EMA calculations for CdSe nanoplatelets by varying the monolayer (ML) thicknesses from 4.5 ML (11.9 Å thick, excluding the ligands) to 5.5 ML (14.9 Å) and to 6.5 ML (18.0 Å). Here, we fixed the latteral dimensions at 20 nm × 20 nm such that the later-direction quantum confinment effect is negligible. In figure 6(c), we present the calcualted quasiparticle gap $E_g^{qp}$ (blue squares) and optical gap $E_g^{opt}$ (blue empty squares) values. Comparing with the expeirmental $E_g^{opt}$ data [14-17], we then find excellent agreements, reaffirming the reliability and generality of our multiscale approach.

#### 4.2.2 Impacts of the bulk-based EMA parameters and abrupt confinement potential

Having accurate EMA parameters and confinement potential shape, it would be informative to systematically asses the effects of individual factors out of the bulk $m_e^* / m_h^*$, bulk $\epsilon$, and abrupt confinement potential approximations. In table 3, we sumarized for the 4.5 ML CdSe nanoplatelet case how $E_g^{qp}$, $E_X$, and $E_g^{opt}$ values are modified by such individual changes. Then, we find that while $E_g^{qp}$ is most significantly affected by the potential shape, $E_X$ is more strongly affected by the effective masses and dielectric constant. In terms of the combined effect for $E_g^{opt}$ [equation (4)], the abrupt potential shape turns out to be the most important source of error. This will be a useful guideline for future computational studies of quantum nanostructures.

## 5. Conclusions and Outlook

In summary, we developed a first-principles based EMA calculation approach for quantum nanostructures, and implemented the method within our grid-based OORE framework (OOREQD) [7-12]. The essential ingredient in the developed scheme is carrying out DFT calculations for reference nanostructures (nanowires for nanorods and nanosheets for nanoplatelets) to (1) extract the nanoscopic effective mass and dielectric constant information and to (2) generate from the atomistic KS potential a realistic EMA confinement potential. Given that the size and shape of the confinement potential are important factors in determining the electronic and optical properties of nanostructures [5, 6], the ability to accurately extract the confinement potential profile for the efficient yet accurate EMA calculation approach will have important implications for the computational design of semiconductor nanostructures.

We applied the developed method to study the optical properties of 1D CdS/ZnS core/shell quantum rods [13] and 2D CdSe nanoplatelets [14-17]. In both cases, we obtained the optical gap $E_g^{opt}$ values in excellent agreement with experimental data. For the nanorods, we found that the optical gap $E_{gap}^{opt}$ or the PL peak position is essentially determined by the nanorod diameter or the radial direction quantum confinement. On the other hand, the length of nanorods or the axial direction quantum confinement was found to affect the overlap between electron and hole wavefunctions and accordingly the PL intensity. In terms of the methodological aspects, we determined that the abrupt confinement potential approximation would incorrectly describe the quantum confinement effect or the quasiparticle gap $E_g^{qp}$ and will become the most significant source of error.

In closing, we comment on several aspects of the developed formalism that can be extended and/or improved in future studies. First, while we focused on the CBM-to-VBM transitions that will be relevant for LED applications, the method can be further extended to the simulation of absorption spectra by including additional band edge states that have been split by the spin-orbit interaction, crystal field effect, etc. Next, in view of the importance of the dielectric screening in determining the characteristics of excitons [49], we plan to explore the position-dependent dielectric function [49, 50]. Finally, we remind that one of the main motivations to employ non-0D nanostructures for optoelectronic applications is the distinctive electric-field-induced PL switching propery [51, 52]. For this purpose, however, the independent electron/hole approximation adopted in this work would not be sufficient to properly describe the electron and hole pair experiencing the strong electron-hole interaction in addition to the external electric field effect. Namely, a non-perturbative approach would be necessary for the proper simulation of the electric-field-induced PL switching property, and for this purpose we initally tested the iterative Hartree scheme [51, 52]. Unfortunately, however, we determined that the coupled Hartree level deteriorates the results reported in this paper and the additional inclusion of the electron-hole exchange effect is essential. Stuides along these lines will be reported in future publications.

## Acknowledgements

This work was supported by the Basic Research Program (No. 2017R1A2B3009872), Nano-Material Technology Development Program (No. 2016M3A7B4024133), Global Frontier Program (No. 2013M3A6B1078881), and Basic Research Lab Program (No. 2016M3A7B4909944) of the National Research Foundation funded by the Ministry of Science and ICT of Korea. Computational resources were provided by the KISTI Supercomputing Center (KSC–2017-C3-0063).

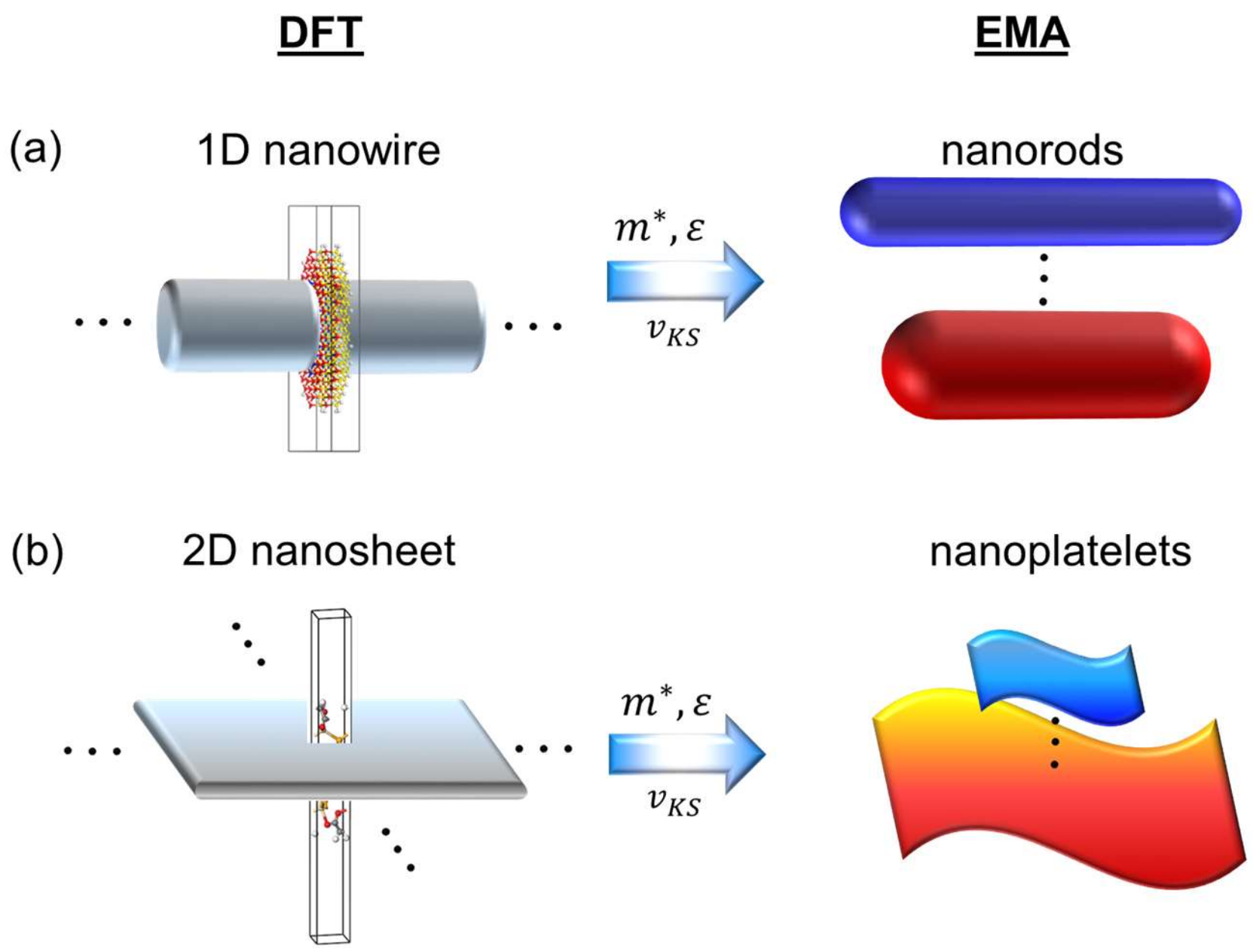

**Figure 1.** Schematics of the proposed DFT-based EMA calculation strategy. For the (a) nanorod and (b) nanoplatelet cases, e.g., we perform DFT calculations for the corresponding ideal 1D nanowire and 2D nanosheet or slab, respectively, and extract the nanoscopic electron/hole mass $\boldsymbol{m}_{e/h}^*$, dielectric constant $\boldsymbol{\epsilon}$, and the confinement potential profile $\boldsymbol{v}^{KS}$ for EMA calculations.

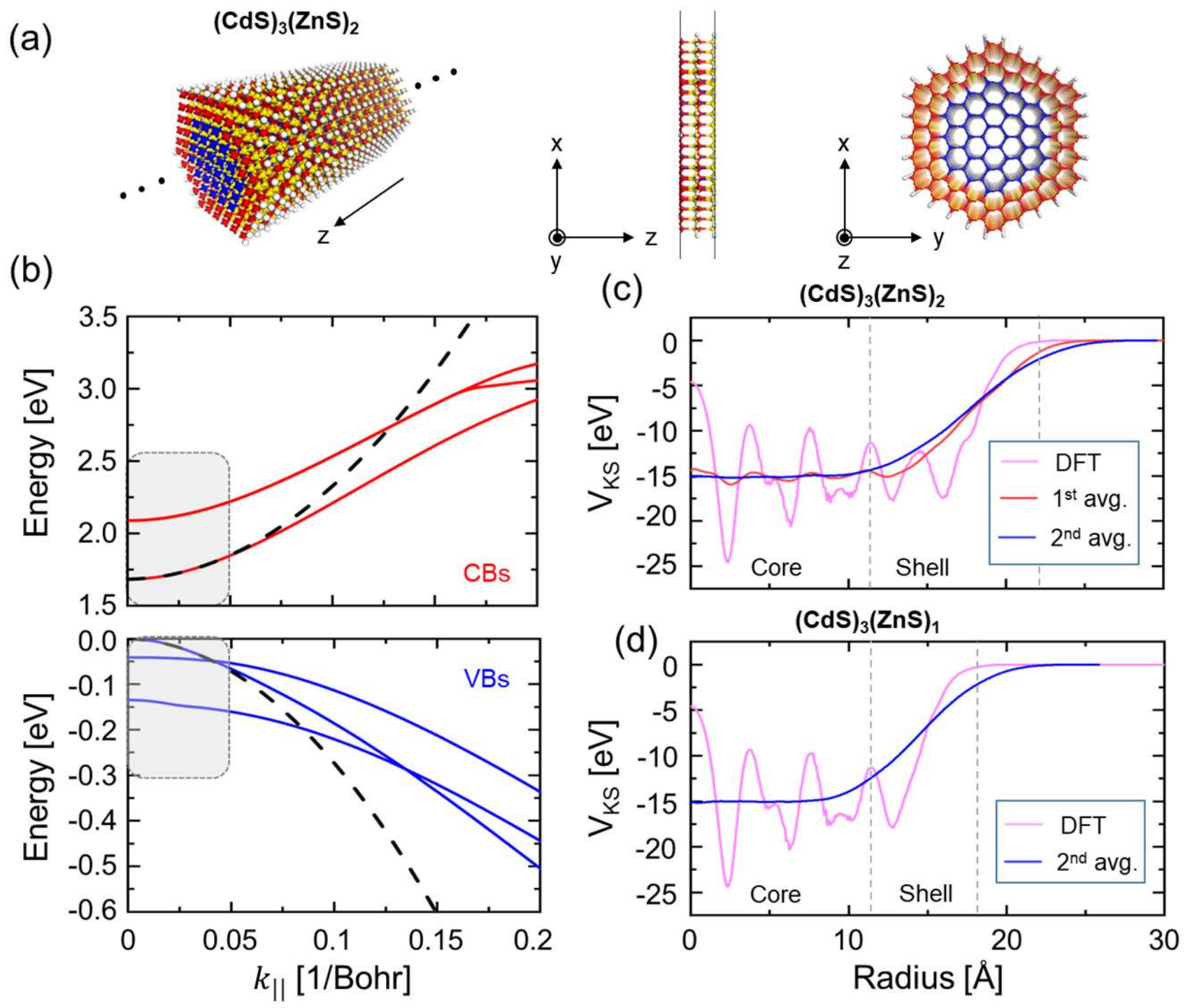

**Figure 2.** (a) Atomistic models of the $(CdS)_3/(ZnS)_2$ core/shell nanowire. (b) Conduction band (top panel) and valance band (bottom panel) structures of the $(CdS)_3/(ZnS)_2$ nanowire calculated by DFT. Effective mass fittings to the DFT-derived CBM and VBM curves are shown as black dotted lines. The shaded region indicates the range of DFT bands utilized for the fittings. (c) From the radially averaged KS-DFT potentials of the $(CdS)_3/(ZnS)_2$ nanowire, a double filtering process generates a smooth confinement potential that can be used for EMA calculations. (d) The radially averaged KS-DFT potential of the $(CdS)_3/(ZnS)_1$ nanowire and the EMA potential obtained through the double filtering process.

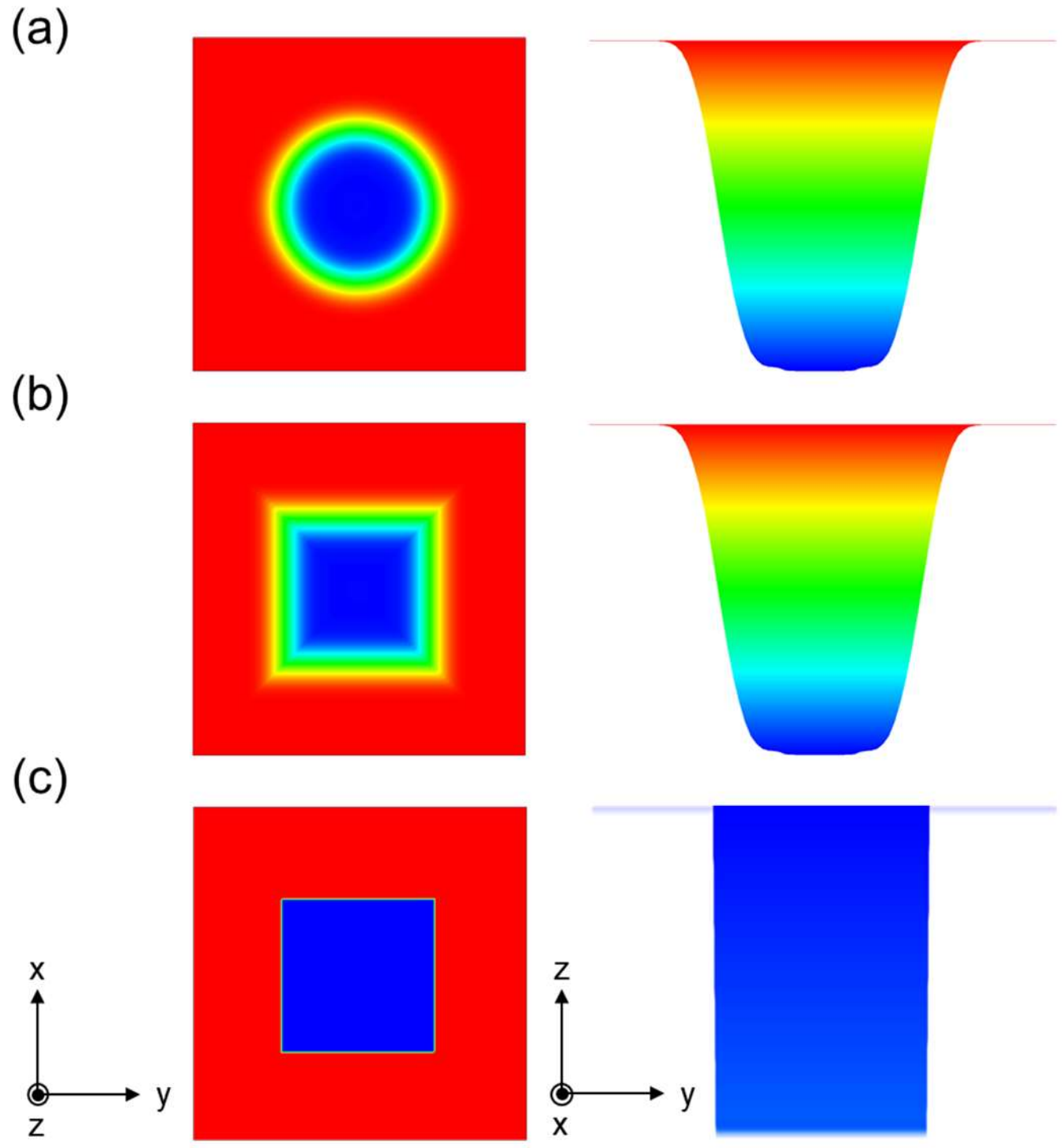

**Figure 3.** Top-view and Side-view of the EMA effective potentials with the shape based on (a,b) DFT KS potential and (c) abrupt step-potential. Radial smoothened DFT KS potentials shown in figure 2(c) were directly converted into (a) cylindrical and (b) rectangular potential shape.

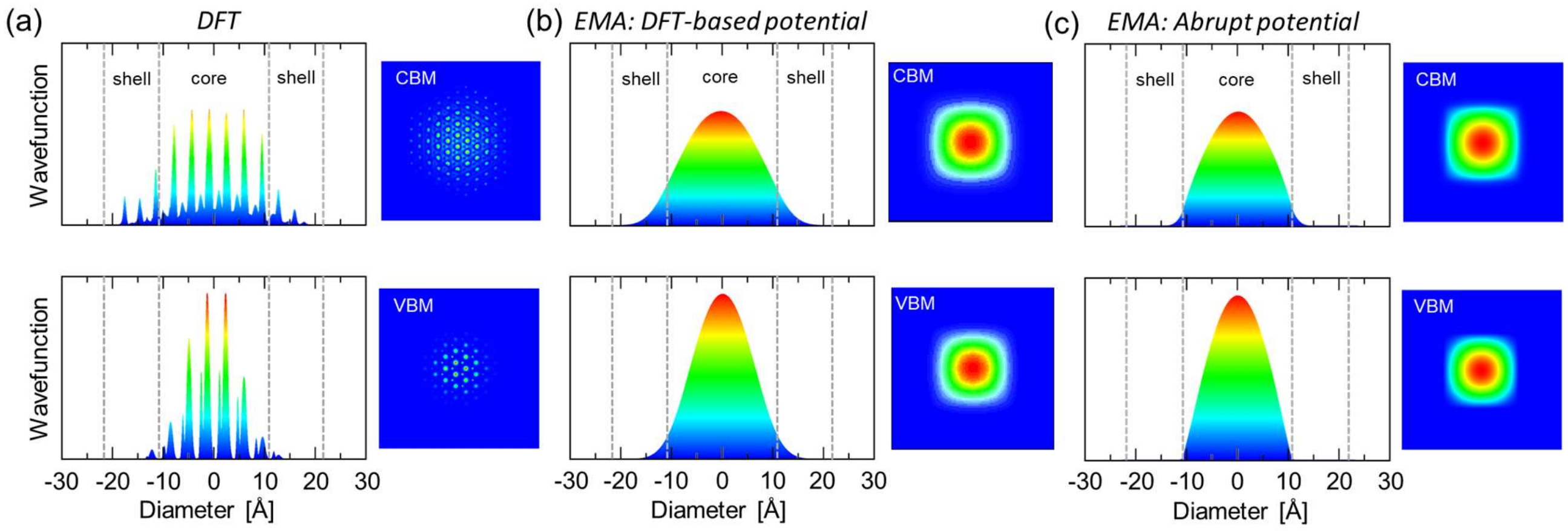

**Figure 4.** The radial cross-sectional view of the electron (top) and hole (bottom) wavefunctions of CdS/ZnS core/shell nanostructures generated within (a) DFT, (b) DFT-KS potential based EMA and (c) abrupt potential based EMA.

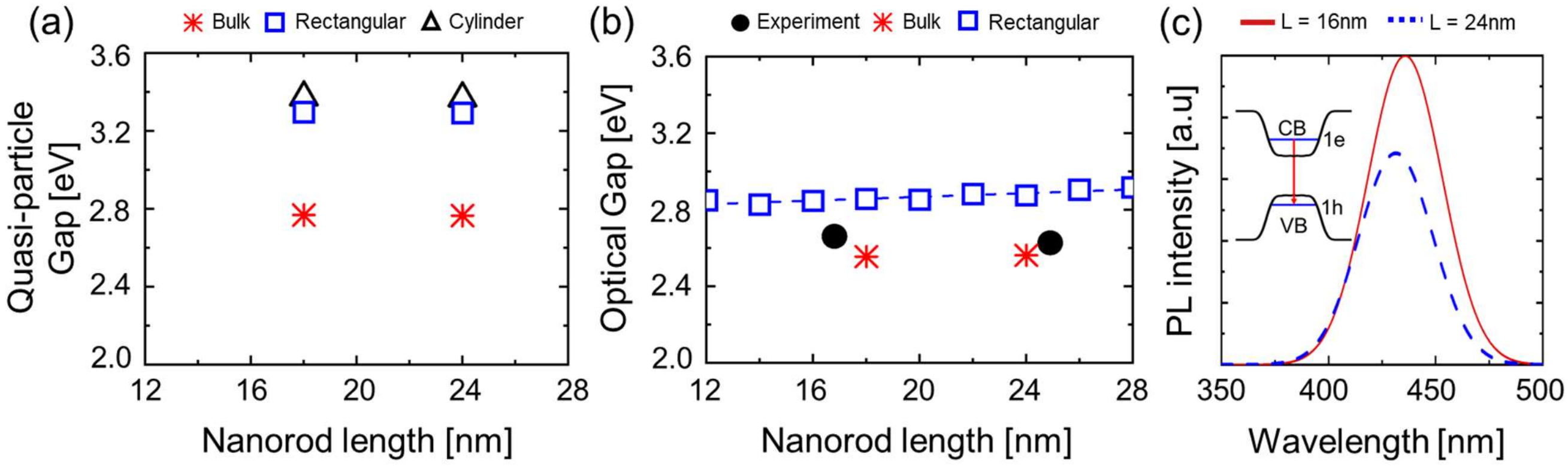

**Figure 5.** (a) Quasiparticle gap $\boldsymbol{E_g^{qp}}$ of $(CdS)_3/(ZnS)_2$ nanorods with varying nanorod lengths calculated by the EMA approaches based on the nanoscopic EMA parameters with cylindrical (black empty triangles), rectangular potential (blue empty squares). $\boldsymbol{E_g^{qp}}$ based on bulk EMA parameters together with an abrupt confinement potential (red stars) is shown together. (b) Comparison of the experimental optical bandgaps $\boldsymbol{E_g^{opt}}$ of $(CdS)_3/(ZnS)_2$ core/shell nanorods (black filled circles) and $\boldsymbol{E_g^{opt}}$ obtained from the DFT-based EMA approaches or the nanoscopic EMA parameters and DFT-based potential (blue empty squares) and $\boldsymbol{E_g^{opt}}$ obtained with the bulk EMA parameters and abrupt potential (red stars). Experimental data are from ref [13]. (c) The PL intensities calculated for $(CdS)_3/(ZnS)_2$ core/shell nanorods with the lengths of 16nm (solid red line) and 24nm (dotted blue line). Inset: A schematic describing the electron transition from the electron to hole energy bands.

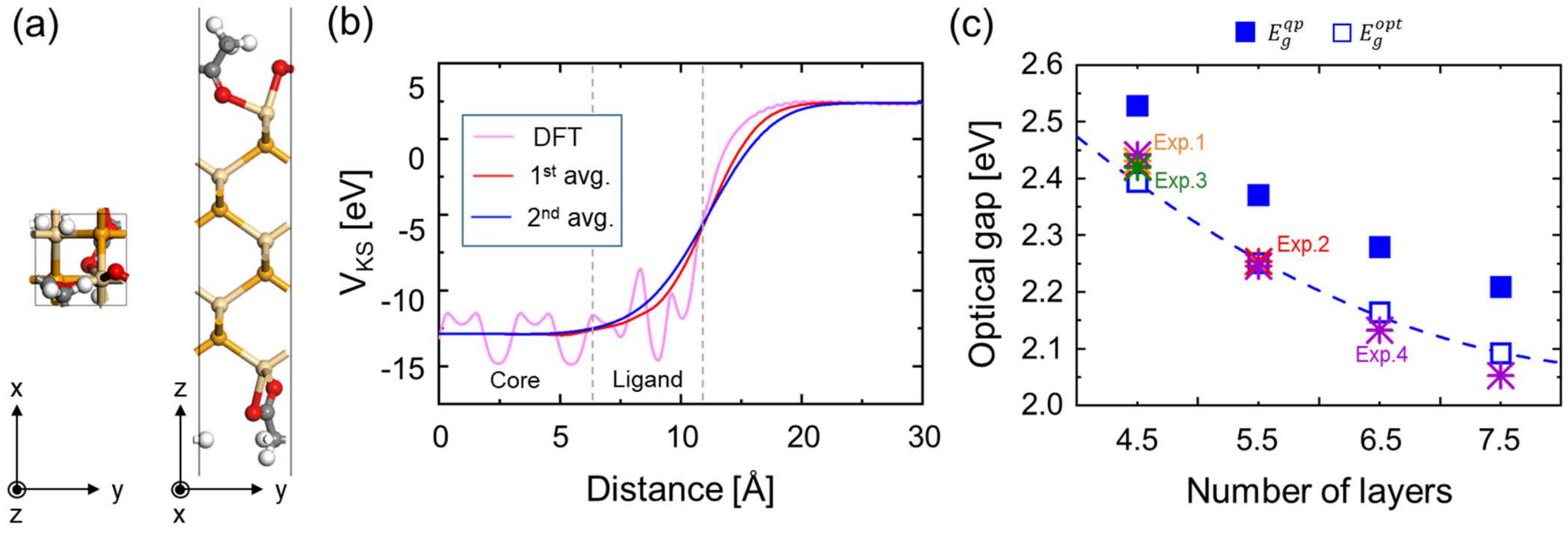

**Figure 6.** (a) Top (left) and side (right) views of the 4.5 ML CdSe nanoplatelet. (b) Plane-averaged KS-DFT potential and sinlge- and double-filtered potentials. (c) Quasiparticle gap $\boldsymbol{E_g^{qp}}$ (blue filled squares) and optical gap $\boldsymbol{E_g^{opt}}$ (blue empty squares) values of the 4.5 ML, 5.5 ML, and 6.5 ML CdSe nanoplatelets calculated by the DFT-based EMA methods. The experimentally reported optical bandgap $\boldsymbol{E_g^{opt}}$ valulues (stars) are shown together. Experimental data are from Ref. [14] (Exp. 1, violet star), Ref. [15] (Exp. 2, orange star), Ref. [16] (Exp. 3, red star), and Ref. [17] (Exp. 4, green star).

| **EMA parameters** | **Symbol (unit)** | **CdS bulk** | **$(CdS)_3(ZnS)_2$ nanowire** |
|---|---|---|---|
| Effective electron mass | $m_e^*$ $(m_0)$ | 0.20 | 0.20 |
| Effective hole mass | $m_\mathrm{h}^*$ $(m_0])$ | 0.70 | 0.51 |
| Dielectric constant | $\varepsilon_r$ | 8.92 | 2.30 |
| Length scaling factor: electron | $a_e^*$ (Å]) | 23.59 | 6.08 |
| Length scaling factor: hole | $a_\mathrm{h}^*$ (Å) | 6.74 | 2.39 |
| Energy scaling factor: electron | $Ry_e^*$ (eV]) | 0.068 | 1.029 |
| Energy scaling factor: hole | $Ry_\mathrm{h}^*$ (eV) | 0.239 | 2.623 |

**Table 1.** EMA parameters derived from bulk CdS and from CdS/ZnS nanowires.

| **Parameter** | **Symbol (unit)** | **CdSe bulk** | **CdSe nanosheet (4.5 ML)** | **CdSe nanosheet (6.5 ML)** |
|---|---|---|---|---|
| Effective electron mass | $m_e^*$ $(m_0)$ | 0.06 | 0.23 | 0.14 |
| Effective hole mass | $m_\mathrm{h}^*$ $(m_0])$ | 0.64 | 4.73 | 3.91 |
| Dielectric constant | $\varepsilon_r$ | 6.20 | 3.10 | 3.54 |
| Length scaling factor: electron | $a_e^*$ (Å) | 10.67 | 20.57 | 13.38 |
| Length scaling factor: hole | $a_\mathrm{h}^*$ (Å) | 103.33 | 13.48 | 0.48 |
| Energy scaling factor: electron | $Ry_e^*$ (eV]) | 0.002 | 0.02 | 0.304 |
| Energy scaling factor: hole | $Ry_\mathrm{h}^*$ (eV) | 0.02 | 0.49 | 8.490 |

**Table 2 .** EMA parameters derived from bulk CdSe and from 4.5 ML and 6.5 ML CdS platelets.

| Value | Symbol (unit) | Nano parameter | Bulk $m_{e/h}^*$ | Bulk $\varepsilon_r$ | Abrupt potential |
|---|---|---|---|---|---|
| Quasi particle gap | $E_g^{qp}$ (eV) | 2.528 | 2.811 | 2.528 | 2.099 |
| Exciton binding energy | $E_x$ $(eV)$ | 0.134 | 0.067 | 0.068 | 0.116 |
| Optical gap | $E_g^{opt}$ (eV) | 2.395 | 2.744 | 2.460 | 1.983 |

**Table 3.** For the 4.5 ML CdSe nanoplatelet case, the impacts of using the bulk effective mass, bulk dielectric constant, and abrupt confinement potential approximations are separately estimated against the DFT-based results.